\documentclass[osajnl,twocolumn,showpacs,superscriptaddress,10pt]{revtex4-1}

\usepackage{amsmath,amssymb,graphicx}

\begin{document}

\title{Spectral Engineering with Coupled Microcavities:\\ Active Control of Resonant Mode-Splitting}

\author{Mario C. M. M. Souza}
\author{Guilherme Rezende}
\address{Instituto de Fisica Gleb Wataghin, Universidade Estadual de Campinas, 13083-970 Campinas, SP, Brazil}
\author{Luis A. M. Barea}
\address{Instituto de Fisica Gleb Wataghin, Universidade Estadual de Campinas, 13083-970 Campinas, SP, Brazil}
\address{Departamento de Engenharia Eletrica, Universidade Federal de Sao Carlos, 13565-905 Sao Carlos, SP, Brazil}
\author{Antonio A. G. von Zuben}
\author{Gustavo S. Wiederhecker}
\author{Newton C. Frateschi}\email{fratesch@ifi.unicamp.br}
\address{Instituto de Fisica Gleb Wataghin, Universidade Estadual de Campinas, 13083-970 Campinas, SP, Brazil}

\begin{abstract}
Optical mode-splitting is an efficient tool to shape and fine-tune the spectral response of resonant nanophotonic devices. The active control of mode-splitting, however, is either small or accompanied by undesired resonance shifts, often much larger than the resonance-splitting. We report a control mechanism that enables reconfigurable and widely tunable mode-splitting while efficiently mitigating undesired resonance shifts. This is achieved by actively controlling the excitation of counter-traveling modes in coupled resonators. The transition from a large splitting (80 GHz) to a single-notch resonance is demonstrated using low power microheaters (35 mW). We show that the spurious resonance-shift in our device is only limited by thermal crosstalk and resonance-shift-free splitting control may be achieved. 
\end{abstract}

\ocis{(230.4555) Coupled resonators; (130.3120) Integrated optics devices.}

\maketitle %% required

The ability to tailor the spectral features of compact photonic devices is essential to address the need for complex optical responses in a wide range of applications, including optical signal processing, sensing and nonlinear optics~\cite{willner_all-optical_2014,claes_label-free_2009,kippenberg_microresonator-based_2011}.
Microcavity resonant mode-splitting has been explored in these applications as a powerful tool to enable spectral engineering with low power consumption and small footprint. 
For example, selective mode-splitting has recently enabled four-wave mixing through dispersion compensation~\cite{gentry_tunable_2014,lu_selective_2014}, carrier recycling~\cite{barea_silicon_2013}, compact phase-shifters for microwave photonics~\cite{chang_tunable_2009}, and the use of multiple-split resonances to overcome the trade-off between the cavity free spectral range (FSR) and optical field enhancement \cite{souza_embedded_2014}.
The engineering of the optical phase delay through resonant mode-splitting has also been explored to demonstrate pulse delay/advancement~\cite{li_fast_2009}.

A major limitation of spectral engineering through mode-splitting in integrated devices is its challenging control. 
In order to be useful for any dynamic implementation one should be able to actively control the cavity mode-splitting without any undesirable global shift of the split resonance.
So far the demonstrated controlling approaches are limited with respect to active tunability or resonance shift. 
For example, counter-propagating mode-splitting induced by controlled sidewall gratings enables large and predictable mode-splitting~\cite{lu_selective_2014}, but it is not tunable.
Tunable mode-splitting has been demonstrated using Mach-Zehnder interferometers to control the coupling strength between degenerate traveling-wave resonators ~\cite{atabaki_tuning_2010,sun_investigation_2013} with high power efficiency, but at the price of resonance shifts of the order of several FSRs.
In practice, any tuning mechanism acting upon an optical path where the resonant optical field is highly confined will necessarily cause a significant change in the optical phase and lead to an undesirable resonance shift. 

In this Letter we demonstrate the active control of mode-splitting in a multi-GHz range with very-low resonance shift in a coupled-cavity device. The large splitting and small resonance-shift are simultaneously achieved through the controlled excitation of counter-traveling modes using embedded microrings.  Our approach effectively separates the tuning cavity section from the target resonant optical mode. We compare our results with theoretical calculations and we show that the performance of our fabricated device is only limited by thermal crosstalk.

\begin{figure}[htb]
\centerline{\includegraphics[width=\columnwidth]{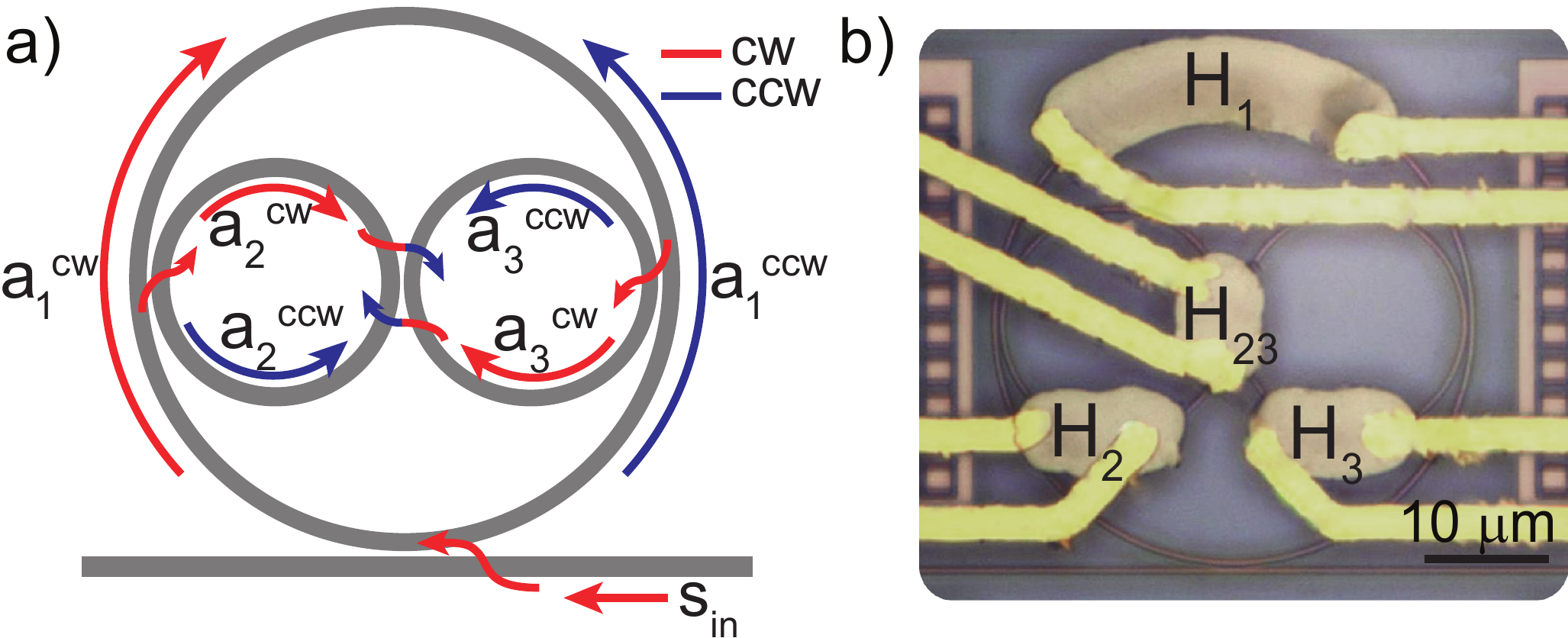}}
\caption{Coupled microring device. (a) The direct coupling between two embedded microrings enables the excitation of CW- and CCW-traveling modes $a_2^{\text{cw,ccw}}$ and $a_3^{\text{cw,ccw}}$ even when the incident light $s_{in}$ travels in one direction only. The coupling between $a_1^{\text{cw}}$ and $a_1^{\text{ccw}}$ is strong only when the detuning between the outer and embedded rings is small. (b) Optical micrography of the fabricated coupled-cavity device with Ni-Cr microheaters. $H_1$, $H_2$ and $H_3$ tune each microring separately, while $H_{23}$ allows to tune both embedded microrings simultaneously.}
\end{figure}

We use a coupled microring device (Fig. 1) that allows for the controlled excitation of clockwise (CW) and counterclockwise (CCW) modes ~\cite{souza_embedded_2014,poon_wavelength-selective_2004}. 
When the light launched into the bus waveguide ($s_{in}$) is resonant with the embedded microrings ($R_2$ and $R_3$), both CW and CCW modes $a_2^{\text{cw,ccw}}$ and $a_3^{\text{cw,ccw}}$ are excited due to the direct coupling between these rings in the central region (Fig. 1a). When the incident light is resonant with the outer microring ($R_1$) only, it excites the CW mode $a_1^{\text{cw}}$. Neglecting CW-CCW coupling through sidewall roughness, the CCW mode $a_1^{\text{ccw}}$ can be excited in the outer ring only if the detuning between embedded and outer microrings is small. $R_2$ and $R_3$ work as resonant back-reflectors and the CW-CCW coupling induced mode-splitting can be dynamically controlled by acting on these microrings. Here we demonstrate this principle using the thermo-optic effect through integrated microheaters (Fig. 1b).
 
\begin{figure}[b]
\centerline{\includegraphics[width=\columnwidth]{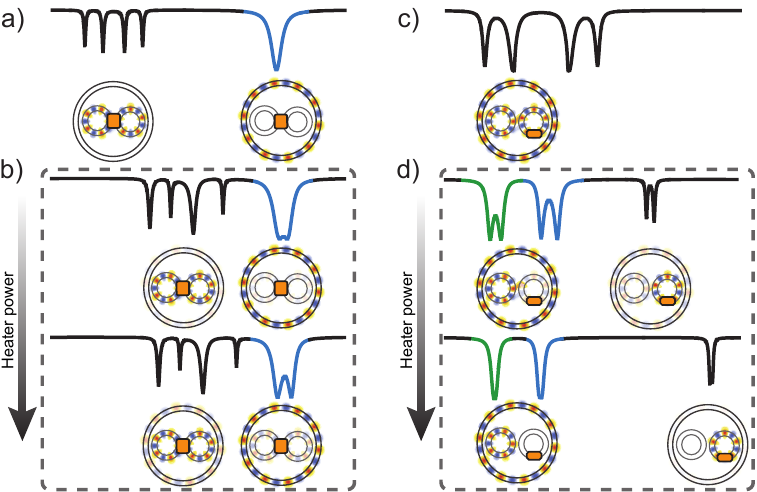}}
\caption{Schematic of the mode-splitting control approaches using microheaters. The heaters are shown in orange and the resonances in which mode-splitting is controlled are shown in blue and green. \textbf{a)} Embedded and outer ring supermodes are initially uncoupled. \textbf{b)} The embedded ring resonances are simultaneously shifted towards the outer ring resonance by means of heater $H_{23}$, increasing the CW-CCW mode-coupling in the outer ring. \textbf{c)} Starting with the three microrings degenerate, \textbf{d)} a single embedded ring is detuned using heater $H_3$, reducing the CW-CCW-induced mode-coupling for all supermodes.}
\label{fig:scheme}
\end{figure}

We control the CW-CCW mode-splitting using two approaches that significantly reduce undesired resonance shifts. This scheme is illustrated in Fig. \ref{fig:scheme} and relies on actuating either both embedded microrings simultaneously (Fig. \ref{fig:scheme}a,b), or only one of them (Fig. \ref{fig:scheme}c,d). 
When $R_2$ and $R_3$ are initially blue-detuned with respect to $R_1$ (Fig. 2a), heater $H_{23}$ red-shifts the four-fold resonances of $R_2$ and $R_3$ towards the single-notch resonance of $R_1$ (blue trace). A controllable CW-CCW mode-splitting is created as mode $a_1^{\text{ccw}}$ is increasingly excited through the embedded rings, as shown in the blue-highlighted trace in Fig. 2b. Alternatively, starting with the three coupled rings degenerate (Fig. 2c), all the six CW and CCW modes ($a_{1,2,3}^{\text{cw,ccw}}$) are coupled but only four resonance notches appear due to an accidental degeneracy ~\cite{souza_embedded_2014}. Using heater $H_3$ to red-shift only $R_3$ effectively reduces the CW-CCW coupling for all supermodes (Fig. 2d). Although the actuated ring supermodes strongly red-shift, the position of the remaining CW-CCW supermodes of cavities $R_{1,2}$ (blue and green traces) are practically  unaffected. 

The proposed splitting control approaches feature important advantages towards widely tunable splitting with no resonance-shift.
The controlled CW-CCW mode coupling is mediated by the coupling between microrings, therefore the maximum CW-CCW induced splitting can be tailored through cavity design (gap separation and coupling length). On the other hand the controlled mode-splitting can always be reduced to a single-notch resonance, since the uncoupled CW and CCW modes are degenerate.
In both schemes described in Fig. 2 the resonance shifts are effectively mitigated as we minimize the overlap between the spatial distribution of the controlled supermodes (blue and green traces) and the cavity sections in which the refractive index is changed (around the orange heaters).

The passive optical device was fabricated at IMEC-EUROPRACTICE with microring radius $R_1$ = 20 $\mu$m, $R_2$ = $R_3$ = 9.625 $\mu$m, and a 200 nm coupling gap both between each microring and between the outer microring and the bus waveguide. Microheaters and contact pads were fabricated in a post-process step using a Ni-Cr and Ti/Au deposition (100 nm and 2/200 nm respectively) and a photoresist lift-off process ~\cite{fegadolli_reconfigurable_2012}.
The measured electrical resistance of microheaters ($H_1$, $H_2$, $H_3$, $H_{23}$) were (275, 99, 107, 104) ohm. Throughout the measurements the voltage applied to the microheaters was kept below 2 V, compatible with CMOS voltage standards.

\begin{figure*}[tp]
\centerline{\includegraphics[width=2\columnwidth]{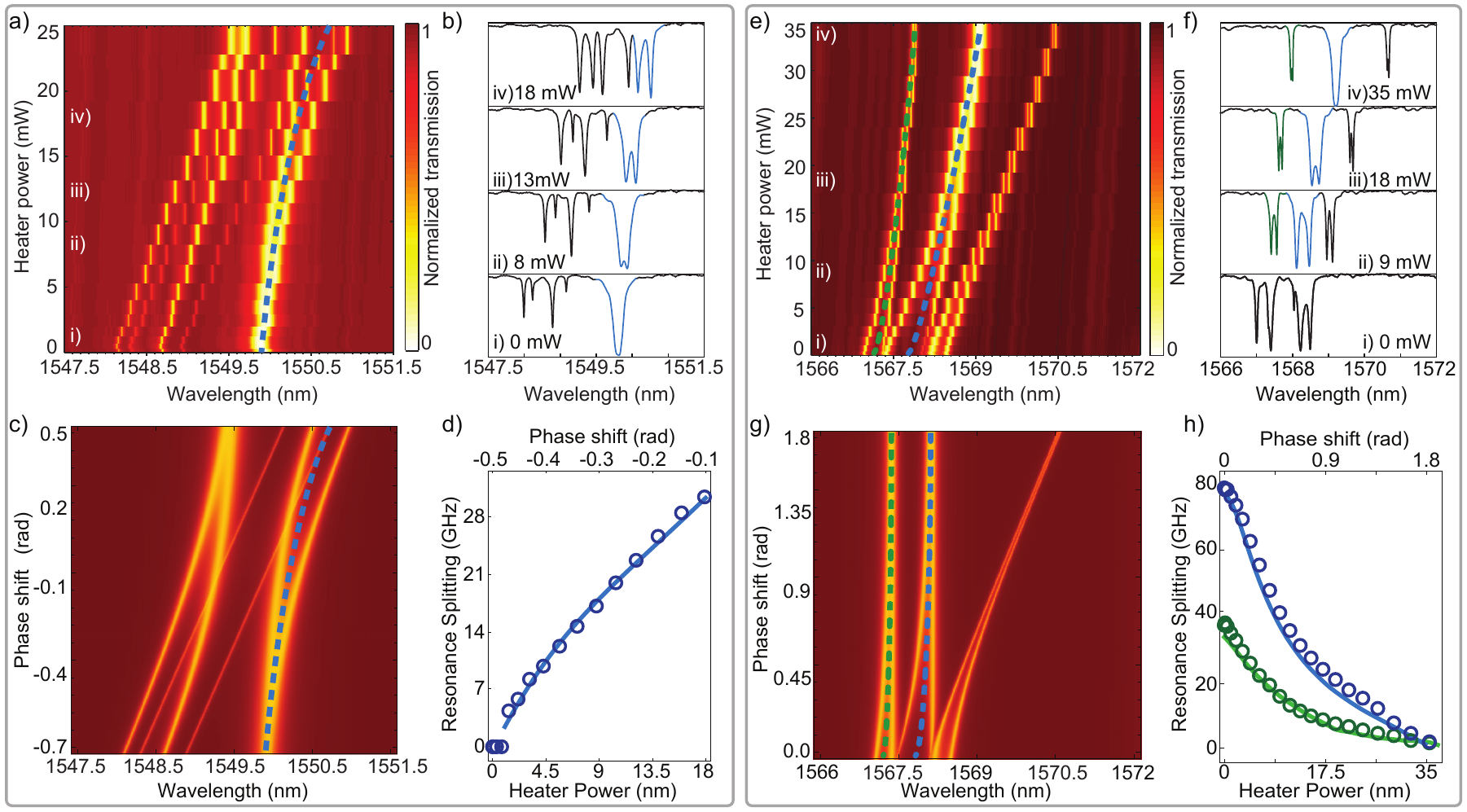}}
\caption{Experimental and calculated spectral evolution of the coupled-cavity supermodes using heaters $H_{23}$ (a-d) and $H_3$ (e-h). \textbf{a,c)} Measured (a) and calculated (c) optical transmission as the heater $H_{23}$ power is increased. The bue-dashed line follows the central wavelength of the controlled resonance. \textbf{b)} Transmission traces corresponding to labelled heater powers. The blue color highlights the controlled resonance. \textbf{c)} Calculated transmission as the phase of the embedded rings are simultaneously varied. \textbf{d)} Experimental (circles) and calculated (line) resonance-splitting as a function of the power dissipated in heater $H_{23}$. Equivalent phase shift between embedded and outer rings is shown in the upper horizontal scale. \textbf{e,g)} Measured (e) and calculated (g) optical transmission as the heater $H_{3}$ power is increased. The blue and green-dashed lines follows the central wavelength of the controlled resonance. \textbf{f)} Transmission traces corresponding to labelled heater powers. The blue and green color highlights the controlled resonances. \textbf{h)} Measured (circles) and calculated (lines) resonance-splitting for the set of split resonances. Green and blue indicate the leftmost and central resonances respectively.}
\end{figure*}

The experimental realization of the tuning approach using heater $H_{23}$ demonstrates the generation and continuous tuning of mode-splitting (Fig. 3a,b). Starting from a single-notch resonance (loaded-Q of 6,700), the mode-splitting emerges at low heating power (7 mW) and increases almost linearly at a rate of roughly 1.8 GHz/mW (Fig. 3d) as the quadruplet resonances are red-shifted torwards the outer ring resonance. From trace (iv) in Fig. 3b, we infer a resonance-splitting of approximately 30 GHz at 17.7 mW of heating power; further heating significantly changes the desired double-notch resonance-splitting. A very small resonance-shift is observed in the transmission spectra (blue-dashed line) due to the mode anti-crossing dispersion. This heating configuration shifts the embedded microring resonances at a rate of 9.5 GHz/mW while a spurious shift of the outer ring resonance occurs at a rate of 1.5 GHz/mW due to a small thermal crosstalk.  To accurately measure the spurious shift we used a spectral region (not shown in Fig. 3) where the embedded and outer rings resonances are uncoupled throughout the full range of heating power.

A qualitatively distinct tunable mode-splitting spectrum is observed when the three microrings are initially degenerate and only one embedded ring ($R_3$) is actively detuned (Fig. 3e,f). At low heating powers, the six supermodes exhibit a quasi-fourfold mode-splitting, highlighted in the transmission trace (i) of Fig. 3f. As $R_3$ is detuned, the coupling strength between CW and CCW-modes gradually vanishes along with the associated mode-splitting for all supermodes. The two resonances of interest are shown in Fig. 3f in green and blue traces. We show the resonance-splitting dependence on the heating power for these resonances in Fig. 3h. Although this dependence is clearly nonlinear this approach yields larger tunable splitting when compared to the previous one, with a maximum splitting of 80 GHz observed for the central (blue) resonance when the heater is off. On the other hand, this approach is also more severely affected by thermal crosstalk as indicated by the dashed lines in Fig. 3e. Heater $H_3$ shifts $R_3$ at 8.7 GHz/mW, while thermal crosstalk impinges resonance shifts of 1.56 GHz/mW in $R_2$ and 4.6 GHz/mW in $R_1$.

We carried out calculations based on the Transfer Matrix Method (TMM) to reproduce the supermode spectrum evolution and asses the impact of mode anti-crossing and thermal crosstalk on undesired resonance-shifts. By neglecting thermal crosstalk in the TMM, we can identify the shifts caused only by mode anti-crossing dispersion.  The TMM plot shown in Fig. 3c was obtained by simultaneously varying the two embedded rings roundtrip phases ($\phi_2$ and $\phi_3$, see appendix A) and agrees very well with the experimental data of Fig. 3a. Both the overall position of the controlled resonance (dashed lines) and the splitting evolution (Fig. 3d, solid line) are well reproduced. This good agreement indicates that the small resonance shift observed in the experimental data is dominated by mode anti-crossing dispersion caused by the repuslion of the outer ring resonance by the embedded ring resonances.

The TMM prediction indicates the great potential for resonance-shift-free, large-splitting control when only one embedded ring is detuned (Fig. 3g,h). The almost dispersionless behaviour of the controlled resonance set is emphasized by the dashed lines in Fig. 3g. It shows that the mode anti-crossing dispersion remains much smaller than the splitting over the full control range. In addition, it confirms that the resonance-shifts experimentally observed are indeed due to thermal crosstalk, which is not a fundamental issue. The effects of crosstalk can be eliminated employing distinct dispersion control mechanisms such as carrier effects in silicon or electro-optic effects in other platforms, which exhibit extremely reduced crosstalk. Moreover these mechanisms would enable low-power, ultrafast mode-splitting modulation.
 
In conclusion, we demonstrated reconfigurable mode-splitting control using a compact and power-efficient coupled-microring device. Experimental results show that resonances can be tuned from single-notches to large mode-splitting (80 GHz) with a minimum of resonance shift (limited by thermal crosstalk) through the controlled excitation of counter-propagating modes. We show that the proposed mechanism has the potential to deliver widely tunable resonance-shift-free mode-splitting. Further demonstrations using carrier effects instead of thermal tuning and diligent cavity designs may provide power-efficient ultrafast devices for applications such as reconfigurable optical filtering, single-sideband modulation and microwave photonics.

The authors acknowledge T.P.M Alegre for fruitful discussion and the Brazilian Nanotechnology National Laboratory (LNNano) during the fabrication process. This work was funded by brazilian funding agencies CNPq, CAPES, FOTONICOM (Grant 08/57857-2) and FAPESP (Grants 2014/04748-2, 2012/17765-7).

\appendix
\section{TMM transfer function}
The TMM curves were calculated using the following power transmission T, obtained using the matrix form of Mason\textsc{\char13}s rule ~\cite{mason_power_1954}:
\begin{equation}
 T = \left| \frac{t_1 + \frac{A_1}{\chi_3}(t_1 A_1 \chi_1 + (1+t_1^2)\chi_2)}{1 + t_1 \frac{A_1}{\chi_3}(t_1 A_1 \chi_1 + 2 \chi_2)}\right|^2
\end{equation}
where $A_i=a_i e^{j \phi_i}$ is the field propagation factor of microring $i$ ($i=1,2,3$), with $a_i$ and $\phi_i=(2\pi/\lambda)L_i n_{eff}(\lambda)$ representing the round-trip attenuation factor and accumulated phase-shift. $L_i$ is the perimeter of microring $i$ and a first order approximation of the effective index is assumed, $n_{eff}(\lambda)\approx n_{eff}(\lambda_0)+(\lambda-\lambda_0)\partial n_{eff}(\lambda_0)/\partial \lambda$. The $\chi_i$ coefficients are given by
\begin{eqnarray*}
\chi_1=(((A_2-t_2 t_{23})(A_3-t_3 t_{23})+t_2 t_3 \kappa_{23}^2))^2
\\
\chi_2=\sqrt{t_2 t_3} t_{23} (1+ t_2 t_3 )(A_2 + A_3 )(A_2 A_3+1)+\\
A_2 A_3(t_2^2 t_3^2+1)(1-2 t_{23}^2 )-t_2 t_3 (t_{23}^2 (A_2-A_3 )^2 +\\
A_2 A_3 (A_2 A_3+4)+1)\ \ \ \ \ \ \ \ \ \ 
\\
\chi_3=(((A_2 t_2-t_{23})(A_3 t_3-t_{23})+\kappa_{23}^2))^2 \ \ \ \ 
\end{eqnarray*}

where the field transmission and coupling coefficients ($t_i, \kappa_i$) respect the lossless coupling condition $t_i^2+\kappa_i^2=1$ and represent the coupling between outer ring and bus waveguide ($t_1, \kappa_1$), embedded rings and outer ring ($t_2, \kappa_2, t_3, \kappa_3$), and between embedded rings ($t_{23}, \kappa_{23}$). The attenuation factors ($a_1, a_2, a_3$) and coupling coefficients ($\kappa_1,\kappa_2,\kappa_3,\kappa_{23}$) used in Fig. 3c,g were (0.965, 0.994, 0.994) and (0.47, 0.345, 0.345, 0.2), respectively. The parameters used to calculate the accumulated phase in Fig. 3c were $n_{eff}(\lambda_0)=2.306$, $\partial n_{eff}(\lambda_0)/\partial \lambda=-1.22 \times 10^{-3}\ \text{nm}^{-1}$, $\lambda_0=1549.5\ \text{nm}$, and in Fig. 3g $n_{eff}(\lambda_0)=2.295$, $\partial n_{eff}(\lambda_0)/\partial \lambda=-1.22 \times 10^{-3}\ \text{nm}^{-1}$, $\lambda_0=1568\ \text{nm}$. An additional term $\delta \phi$ was introduced in $\phi_2$ and $\phi_3$ in Fig. 3c and to $\phi_3$ in Fig. 3g accounting for the phase shift in the theoretical plots. The parameters used in the TMM plots were obtained by adjusting the calculated transmission traces to the and experimental ones around the spectral region of interest.

\end{document}